\documentclass[sigconf]{acmart}

\usepackage{booktabs} % For formal tables
\usepackage{tabularx}
\usepackage{tabulary}
\usepackage{listings}
\usepackage{multirow}
\usepackage{subfig}
\usepackage{xcolor}
\definecolor{commentgreen}{RGB}{2,112,10}
\definecolor{eminence}{RGB}{108,48,130}
\definecolor{weborange}{RGB}{255,165,0}
\definecolor{frenchplum}{RGB}{129,20,83}

\setcopyright{rightsretained}

\acmDOI{}

\acmISBN{}

\acmConference[PCSC2024]{Philippine Computing Science Congress}{May 2024}{Laguna, Philippines}
\acmYear{2024}
\copyrightyear{2024}

\acmArticle{}
\acmPrice{}

\editor{}
\editor{}
\editor{}

\begin{document}
\title{AniFrame: A Programming Language for 2D Drawing and Frame-Based Animation}

% Remove for camera-ready version
% \author{Anonymous Author(s)}
% \affiliation{}
% \email{}

\author{Mark Edward M. Gonzales}
\affiliation{%
  \institution{De La Salle University}
  \city{Manila, Philippines}
  \postcode{43017-6221}
}
\email{mark\_gonzales@dlsu.edu.ph}

\author{Hans Oswald A. Ibrahim}
\affiliation{%
  \institution{De La Salle University}
  \city{Manila, Philippines}
  \postcode{43017-6221}
}
\email{hans\_oswald\_ibrahim@dlsu.edu.ph}

\author{Elyssia Barrie H. Ong}
\affiliation{%
  \institution{De La Salle University}
  \city{Manila, Philippines}
  \postcode{43017-6221}
}
\email{elyssia\_ong@dlsu.edu.ph}

\author{Ryan Austin Fernandez}
\affiliation{%
  \institution{De La Salle University}
  \city{Manila, Philippines}
  \postcode{43017-6221}
}
\email{ryan.fernandez@dlsu.edu.ph}

\renewcommand{\shortauthors}{Gonzales, Ibrahim, Ong \& Fernandez}

\begin{abstract}
Creative coding is an experimentation-heavy activity that requires translating high-level visual ideas into code. However, most languages and libraries for creative coding may not be adequately intuitive for beginners. In this paper, we present AniFrame, a domain-specific language for drawing and animation. Designed for novice programmers, it \textit{(i)} features animation-specific data types, operations, and built-in functions to simplify the creation and animation of composite objects, \textit{(ii)} allows for fine-grained control over animation sequences through explicit specification of the target object and the start and end frames, \textit{(iii)} reduces the learning curve through a Python-like syntax, type inferencing, and a minimal set of control structures and keywords that map closely to their semantic intent, and \textit{(iv)} promotes computational expressivity through support for common mathematical operations, built-in trigonometric functions, and user-defined recursion. Our usability test demonstrates AniFrame's potential to enhance readability and writability for multiple creative coding use cases. AniFrame is open-source, and its implementation and reference are available at \url{https://github.com/memgonzales/aniframe-language}.
\end{abstract}




\maketitle

\section{Introduction}

Recent years have seen an increased interest in creative coding, or programming with an artistic rather than a functional intent, especially in engaging artistic expression and computational thinking among novice programmers~\cite{10.1145/2157136.2157342, 10.1145/2157136.2157214}. It has been described as an expressive and exploratory activity~\cite{10.1145/3586183.3606719}, with creative coders having to translate high-level visual ideas into code and go through multiple rounds of incremental refining (or even switching to another idea altogether) depending on the resulting output.

\begin{figure}[t!]
\includegraphics[width=\linewidth]{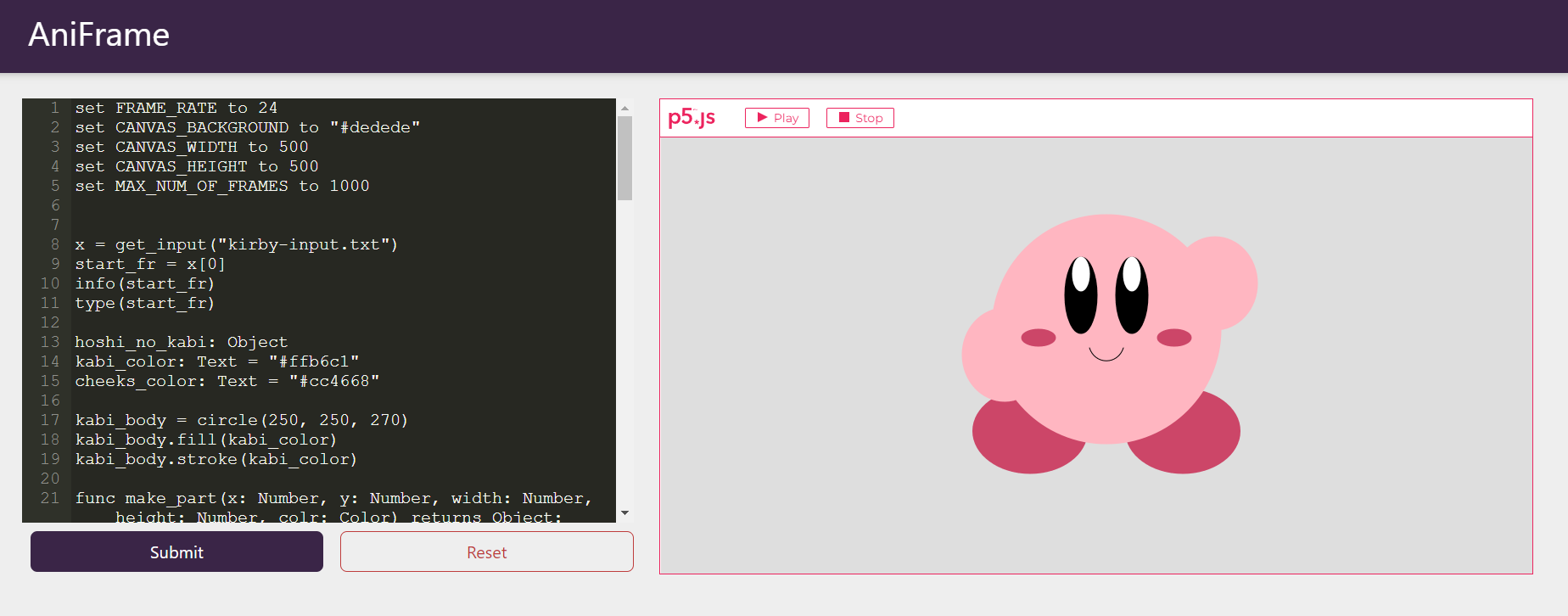}
\caption{Sample AniFrame Code and Resulting Output. \textmd{Designed for novice programmers, AniFrame supports a varied set of features that simplify the creation of composite objects and provide fine-grained control over animation, thus facilitating expressivity and exploration for creative coding.}}
\end{figure}

The difficulty of repeatedly mapping mental models to computer-compatible code is the crux of programming. Hence, from a human-centered viewpoint, programming languages and, by extension, libraries are not only notations or frameworks for writing code but, more importantly, user interfaces purposely designed to facilitate this mapping and make the programming task easier~\cite{10.5555/1978702.1978717}.  

To this end, libraries dedicated to creative coding \cite{p5, cinder, openframeworks, c4, sketch} have been developed to provide an ecosystem and set of constructs for drawing and animation. However, using a library presupposes proficiency in the language for which it was created. Some syntactic and semantic aspects may also not be readily intuitive for novice coders. Examples include p5.js' and Cinder's use of braces for block demarcation, which they carry over from JavaScript and C++, respectively~\cite{10.1145/3077618, p5, cinder}; p5.js' mechanics for layering drawn objects, which have been characterized as challenging~\cite{Sandberg1319386}; and Cinder's interfacing with OpenGL, which can pose a steep learning curve for programmers with no prior experience in graphics APIs~\cite{10.1145/2538862.2538892}. 

Meanwhile, few domain-specific languages for animation have been designed, with most of them, such as ActionScript~3.0~\cite{actionscript} and Processing~\cite{processing}, following an object-oriented programming paradigm. Although this results in highly structured code, having to create classes and instantiate them even for simple programs impedes rapid prototyping and has been noted to increase the difficulty of learning for beginning programmers~\cite{10.1145/3341525.3387369}. 

In this paper, we present AniFrame, an open-source domain-specific language for two-dimensional drawing and frame-based animation for novice programmers. The language's core principles and features are as follows:
\begin{itemize}
    \item \textit{Ready Support for Animation-Specific Constructs.} AniFrame features animation-specific data types (e.g., for drawn objects and colors), operations (e.g., for mixing colors and simplifying the layering of objects into composite objects), and built-in functions for shapes and affine transformations.
    \item \textit{Fine-Grained Control Over Animation.} AniFrame adopts a frame-based strategy where programmers explicitly specify the object to be animated, along with the start and end frames for the animation sequence. Settings such as the frame rate and the total number of frames can also be configured.
    \item \textit{Reduced Learning Curve.} AniFrame follows a Python-like syntax, limits the number of keywords and control structures to a minimum, and tries to use keywords that are close to their semantic intent (e.g., \texttt{Text} instead of \texttt{string}). Specifying data types is optional since type inferencing is enforced.  
    \item \textit{Computational Expressivity.} AniFrame supports common mathematical operations, built-in trigonometric functions, and user-defined recursive functions. Their utility is demonstrated in creating self-similar patterns, such as fractals.
\end{itemize}

\section{Related Works}
% Programming covers a variety of tasks that involve developing a sequence of instructions to be performed by a computer \cite{whatisprog2002}.

% Coding allows users to translate concepts and ideas into specified written instructions and data for a computer to process \cite{drawwithcode2020}. 

% Creative coding in itself is also a form of coding, however it further differentiates itself from traditional programming in that its goal is to express art rather than to create a functional program \cite{Sandberg1319386}. While creative coding, digital experiences of installations and interactive programs can be created, which carry a variety of functions ranging from the recreational \cite{drawwithcode2020} such as simple artistic expression, to the practical \cite{drawwithcode2020} such as introducing coding concepts to non-programmers \cite{drawwithcode2020, Sandberg1319386}. 

% Since interest in creased interest in creative coding coding grew, several tools have been developed to ease the coding process and support creativity among programmers and artists alike \cite{drawwithcode2020}. Sandberg \citeA{Sandberg1319386} broadly classifies these tools into three categories: \textit{(i)}~domain-specific languages, \textit{(ii)}~desktop-based libraries, and \textit{(iii)}~web-based libraries.

Similar to traditional programming, creative coding involves translating ideas into a sequence of instructions for a computer to process. However, it differentiates itself in that its goal is the expression of art rather than the development of functional software~\cite{10.1145/2157136.2157342}. It thus opens a variety of functions, ranging from recreational (e.g.,~making digital sketches and installations~\cite{drawwithcode2020}) to practical (e.g.,~introducing coding concepts to novice and non-programmers~\cite{10.1145/2157136.2157214}) intents. 

With the increased interest in creative coding in recent years, various tools have been developed to ease the coding process and support creativity among programmers and artists alike~\cite{drawwithcode2020}. These tools can be broadly categorized into domain-specific languages, desktop-based libraries, and web-based libraries~\cite{Sandberg1319386}.

Among the few domain-specific languages that have been developed, the most widely used are  Processing~\cite{processing} and ActionScript~\cite{actionscript}. Processing is a simplification of Java for animation. ActionScript was created to accompany the now-discontinued Adobe Flash, and, with its latest release (version 3.0), it became a superset of ECMAScript. Both follow an object-oriented paradigm (OOP), thus promoting reusability and modularity; however, this may be overly complicated for simple drawing and animation use cases, as well as cognitively demanding for novice programmers ~\cite{10.1145/2700519, 10.1145/3341525.3387369}.

Libraries for creative coding were developed later on, as languages and browsers extended support for more features and improved their portability. Desktop-based libraries include vvvv~\cite{vvvv}, which supports more sophisticated functionalities such as 3D rendering and machine learning but runs only on .NET ecosystems; C4~\cite{c4}, which provides a simplified API for mobile-specific features but runs only on iOS; and Cinder~\cite{cinder}, which is cross-platform but can be difficult to learn for those without prior experience with OpenGL~\cite{Sandberg1319386}. Web-based libraries, such as p5.js~\cite{p5} and Sketch.js~\cite{sketch}, are mostly written in JavaScript \cite{Sandberg1319386} and, therefore, presuppose knowledge of JavaScript, which may not be suited for beginners due to its global variable-based programming model~\cite{10.5555/1386753}.

While these tools have proved helpful in a variety of use cases, most domain-specific languages for creative coding assume familiarity with OOP, and libraries require experience with their base ecosystem or language, thereby increasing the learning curve for beginners. AniFrame thus borrows design elements from these tools and also attempts to improve accessibility for novice coders, while still maintaining expressivity for more elaborate use cases.

\section{Language Design}
AniFrame's design is grounded in \textit{(i)}~ready support for animation-specific constructs, \textit{(ii)}~fine-grained control over animation, \textit{(iii)}~reduced learning curve, and \textit{(iv)}~computational expressivity.

\subsection{Animation-Specific Features}
The domain-specificity of AniFrame derives from its animation-specific features, as reflected in its data types, operations, built-in functions, and frame-based approach to animation.

\subsubsection{Data Types and Operations} \label{data-types}
AniFrame features three standard data types: \textit{(i)} \texttt{Text} for strings, \textit{(ii)} \texttt{Number} for floating-point and integer values, and \textit{(iii)} \texttt{List} for collections of data that are possibly heterogeneous (i.e., of different data types). Moreover, it supports three domain-specific types: \textit{(i)} \texttt{Object} for shapes and composite objects, \textit{(ii)} \texttt{Color} for colors (which can be initialized in either hex or RGB), and \textit{(iii)} \texttt{Coord} for coordinate pairs. 

Selected operations between these domain-specific types are also permitted. For example, adding two objects creates a composite object, where the second operand is layered on top of the first operand. Adding two colors is equivalent to color mixing, while subtracting two colors is equivalent to color subtraction. Operations on coordinates are also defined in a component-wise fashion. 

On a syntactic note, the word choices for data types were selected to be as intuitive and close as possible to their semantic intent; for instance, \texttt{Text} was chosen instead of \texttt{string}, and \texttt{Number} instead of \texttt{float} or \texttt{int}, following the results of the empirical study conducted by Stefik and Siebert~\cite{10.1145/2534973} among novice programmers.

\subsubsection{Built-In Functions} \label{funcs}

AniFrame provides a varied set of built-in functions for 2D drawing and animation, as listed in Table~\ref{functions}. 

The largest class of built-in functions comprises those for drawing, styling, and animating objects. To facilitate rapid prototyping, each animation function has a version that applies the affine transformation only on the \(x\)-axis, only on the \(y\)-axis, and on both axes. In terms of syntax, their names purposely depart from their formal mathematical terminologies (e.g., \texttt{move}, \texttt{turn}, and \texttt{resize} instead of \texttt{translate}, \texttt{rotate}, and \texttt{scale}) in order to make their semantic intent more readily understood even by programmers without specialized mathematical background.

Common mathematical functions (e.g., square root and pseudorandom number generation) and trigonometric functions, which are important in programmatic animation~\cite{Peters2006}, are also available out of the box for computational expressivity. 

Moreover, in order to assist in debugging, the built-in method \texttt{info()} can be called to display values or, in the case of objects, their internal representations (discussed in Section~\ref{repr}), while \texttt{type()} can be invoked to display the data types of variables. 

\subsubsection{Frame-Based Animation} \label{frame}
AniFrame adopts a frame-based approach to animation, where \textit{(i)}~all the built-in functions have required parameters for specifying the start and end frames, and \textit{(ii)}~an animation is applied only to the target object calling it. For example, to move an object \texttt{x} by 3 units to the right starting at frame 10 and ending at frame 20, the code is \texttt{x.moveX(3, 10, 20)}. 

This approach affords the programmer fine-grained control over animation sequences while still allowing for rapid prototyping since the appearance and position of the object in the in-between frames are automatically computed under the hood. In this regard, AniFrame takes after the idea of keyframes and tweening in Flash.

AniFrame's principle of applying an animation only to the target object calling it attempts to address pain points in the stack-based approach of OpenGL (and later inherited by Processing and p5.js), where an animation is applied to all the objects created subsequent to the animation function call unless reset via \texttt{pop()}~\cite{opengl-pop, processing, p5}. 

\begin{table*}[t!]
  \caption{Built-In Functions in AniFrame}
  \label{functions}
  \begin{tabulary}{\linewidth}{lL}
    \toprule
    Category & Functions \\
    \midrule 
    Shapes & \texttt{point()}, \texttt{line()}, \texttt{curve()}, \texttt{circle()}, \texttt{ellipse()}, \texttt{triangle()}, \texttt{rectangle()}, \texttt{quad()} (for quadrilaterals), \texttt{polygon()}, \texttt{write()} (for text boxes) \\
    Styling & \texttt{fill()}, \texttt{stroke()} \\
    Translation & \texttt{move()}, \texttt{moveX()}, \texttt{moveY()} \\
    Rotation & \texttt{turn()}, \texttt{turnX()}, \texttt{turnY()} \\
    Scaling & \texttt{resize()}, \texttt{resizeX()}, \texttt{resizeY()} \\
    Shear & \texttt{shear()}, \texttt{shearX()}, \texttt{shearY()} \\
    General Math & \texttt{rand\_num()}, \texttt{rand\_int()}, \texttt{sqrt()} \\
    Trigonometry & \texttt{sin()}, \texttt{cos()}, \texttt{tan()}, \texttt{asin()}, \texttt{acos()}, \texttt{atan()}, \texttt{atan2()}, \texttt{to\_deg()}, \texttt{to\_rad()} \\
    Miscellaneous & \texttt{draw()} (for placing objects on the canvas), \texttt{info()} (for displaying values or, in the case of objects, their internal representations), \texttt{type()} (for displaying data types) \\ 
  \bottomrule
\end{tabulary}
\end{table*}

\subsection{Control Structures}

AniFrame is a Turing-complete language in that it has sequential, conditional, and iterative control structures. With regard to conditionals, two-way selection using \texttt{if...else} and multiple selection using \texttt{else if} are supported. The decision to have only a single multiple selection structure (i.e., not supporting \texttt{switch...case} for instance) is deliberate; AniFrame tries to limit the number of keywords and control structures to a minimum in order to maintain a low learning curve for novice programmers.

AniFrame features three classes of iterative control structures: \textit{(i)}~precondition-controlled loops using \texttt{while}, \textit{(ii)}~count-controlled loops using \texttt{repeat}, and \textit{(iii)}~collection-controlled loops using \linebreak \texttt{for...in}. Including a dedicated count-controlled loop is an attempt to increase readability and writability for use cases such as repeating an object's movement a predefined number of times. A \texttt{break} statement is also provided to prematurely escape loops. However, unconditional branching (e.g., via \texttt{goto}) is not supported in order to discourage "spaghetti" control flows~\cite{10.5555/1241515.1241523}. 

\subsection{Type System and Scoping Rules} \label{scoping}

AniFrame follows a static type system with support for both explicit and implicit type declaration (the latter via type inferencing). The motivation for preferring a static over a dynamic type system is to promote increased reliability and code maintainability~\cite{10.1007/s10664-013-9289-1}. Empirical results~\cite{10.1145/2597008.2597152} have also pointed towards the advantage of static type systems for performing tasks that involve working with previously unknown API functions; in principle, this may also apply to AniFrame since a significant part of the initial learning curve entails developing familiarity with its built-in functions. 

In terms of scoping, AniFrame follows static (lexical) scoping. Creating user-defined functions is supported, but nested functions are not allowed. Any variable declared outside a function is considered a global variable. A function can read the value of a global variable but cannot modify it, except when it is an object; this exception makes it easier to modularize the creation of composite objects by eliminating the need to pass the base object as a parameter (as demonstrated in Section~\ref{comp}). When a local variable inside a function shares the same name as that of a global variable, the global variable cannot be read inside that function. Since the target users of AniFrame are novice programmers, these restrictions are set to simplify the semantics of global scoping and prevent often-unintended side effects when working with global variables~\cite{10.5555/1074100.1074789}.

Syntactically, blocks are demarcated via indentations; in this regard, AniFrame takes after Python, which is well regarded for its readability and writability~\cite{9689813}.

\vspace{4.2mm}

\section{Language Implementation}

AniFrame is an interpreted language. To maintain a strict separation of concerns, its implementation is divided into three stages: \textit{(i)}~lexical analysis, \textit{(ii)}~parsing, and \textit{(iii)}~semantic analysis. The handling modules, namely the lexer (for lexical analysis), parser (for parsing), and interpreter (for semantic analysis), were written in Python using the language recognition tool ANTLR 4~\cite{10.5555/2501720}.

\subsection{Lexical Analysis}

First, the lexer performs lexical analysis, converting the source code into a list of tokens following a longest-match-wins strategy~\cite{10.5555/2501720} and stripping out comments. For illustration, a sample code snippet and its partial token stream are given in Listing~\ref{code} and Table~\ref{lex}, respectively. The patterns for the tokens are defined using regular expressions. In addition, since AniFrame borrows Python's indentation-based block demarcation, special tokens for indentations and outdentations are generated following the stack-based algorithm described in the Python Language Reference~\cite{python-ref}. 

\lstdefinestyle{interfaces}{
  float=tp,
  floatplacement=tp,
  abovecaptionskip=-5pt,
}

\lstset{
    language=Python, 
    commentstyle=\color{black}, keywordstyle=\color{black}, 
    stringstyle=\color{black}, 
    otherkeywords={circle, fill, stroke, ellipse, rectangle, repeat, moveX},
    emph={Object, Color, Coord, List, Text, Number},
    emphstyle={\color{black}},
    classoffset=1, % starting new class
    morekeywords={make_part, make_hand, },
    keywordstyle=\color{black},
    classoffset=0,
    columns=fullflexible, 
    keepspaces=true, 
    breaklines=true, 
    frame=tb, 
    gobble=4, 
    breakindent=0pt, 
    belowcaptionskip=14pt, 
    basicstyle=\ttfamily, 
    mathescape, 
    numbers=left,
    stepnumber=1, 
    xleftmargin=3.3ex
}
\lstset{columns=fullflexible, keepspaces=true, breaklines=true, frame=tb, gobble=4, breakindent=0pt, label={regex}, belowcaptionskip=14pt, basicstyle=\ttfamily, mathescape}
\begin{lstlisting}[style = interfaces, label = code, caption = {
    Sample Code Snippet for Demonstrating AniFrame's Implementation}]
    nose: Object = rectangle(-650, 250, 800, 15) 
    nose.fill("#C2B280")
    frame = 1
    repeat 3:
        nose.moveX(20, frame, frame + 100)
        frame += 101
    pinocchio: Object = circle(0, 250, 200)
    pinocchio += nose
\end{lstlisting}

\begin{table}[t!]
  \caption{Partial Token Stream for Listing~\ref{code}}
  \label{lex}
  \begin{tabularx}{\linewidth}{rrlX}
    \toprule
    Line \# & Column \# & Token & Lexeme \\
    \midrule 
    1 & 1 & \texttt{IDENTIFIER} & \texttt{nose} \\
    1 & 5 & \texttt{COLON} & \texttt{:} \\
    1 & 7 & \texttt{OBJECT\_TYPE} & \texttt{Object} \\
    1 & 14 & \texttt{ASSIGNMENT\_OP} & \texttt{=} \\
    1 & 16 & \texttt{RECTANGLE} & \texttt{rectangle} \\
    1 & 25 & \texttt{OPEN\_PARENTHESIS} & \texttt{(} \\
    1 & 26 & \texttt{MINUS\_OP} & \texttt{-} \\
    1 & 27 & \texttt{INTEGER\_LITERAL} & \texttt{650} \\
    \texttt{...} & \texttt{...} & \texttt{...} & \texttt{...} \\
  \bottomrule
\end{tabularx}
\end{table}

\subsection{Parsing}

\begin{figure}
\includegraphics[width=\linewidth]{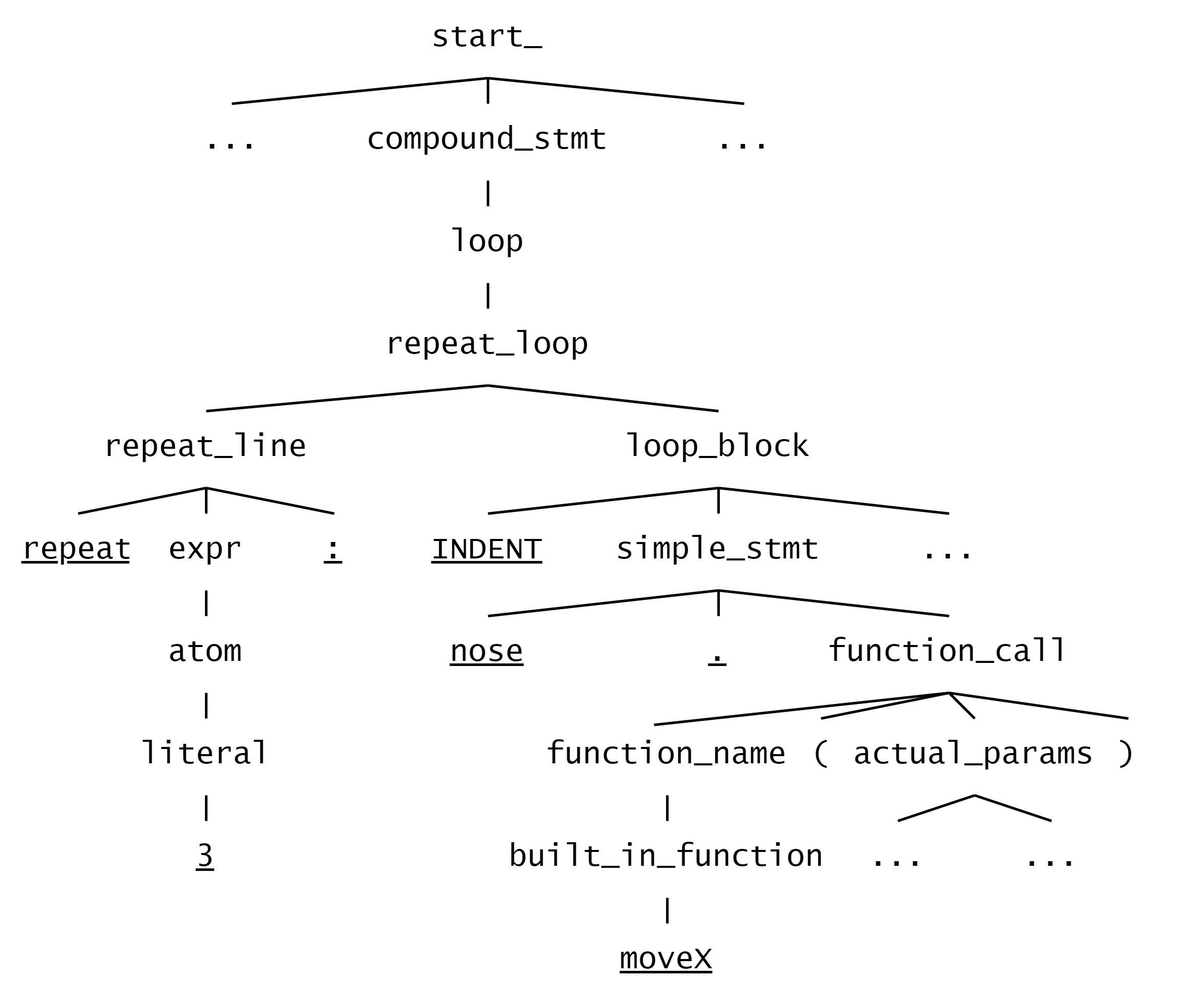}
\caption{Partial Parse Tree for Listing~\ref{code}. \textmd{This figure focuses on the \texttt{repeat} loop (Lines 4 to 6). Terminals are underlined.}}
\label{parse-tree}
\end{figure}

After the lexical analysis, the parser creates a parse tree (Figure~\ref{parse-tree}). This parser is generated via ANTLR's adaptive LL(*) parsing strategy, a predictive approach that involves an arbitrary lookahead and the launching of multiple pseudo-parallel subparsers for efficiency ~\cite{10.1145/2714064.2660202}. AniFrame's grammar is expressed using a variant of extended Backus-Naur form.

\subsection{Semantic Analysis and Code Generation} \label{repr}

After the parse tree is generated, the interpreter traverses it in order to determine the semantic intent of the statements in the source code. This traversal can be performed using either ANTLR's listener or visitor interface~\cite{10.5555/2501720}, but since it is necessary to return the results of visiting certain nodes (e.g., to update the symbol table) and propagate certain values up the parse tree, AniFrame's implementation employs the visitor pattern. 

As the parse tree is traversed and expressions are resolved, the interpreter also builds a JSON-like intermediate code that consists of two dictionaries: \textsc{Drawing} and \textsc{Animation} (Listing~\ref{intermediate-code}). \textsc{Drawing} stores the drawn objects, their constituent shapes, and the styles (i.e., their stroke and fill colors) of each constituent shape. \textsc{Animation} stores the drawn objects, the animations applied to them, and the start and end frames for each animation.

The rationale for this internal representation is twofold. First, although programmers do not have to be aware of this under-the-hood representation, AniFrame provides a function to view it to facilitate debugging, especially for more experienced users. Hence, a conceptually intuitive and human-readable representation is preferable. Second, since the \textsc{Drawing} and \textsc{Animation} dictionaries are updated as nodes in the parse tree are visited, utilizing a data structure that supports \(O(1)\) lookups and updates is critical to speeding up the intermediate code generation.

After the entire parse tree is traversed and the intermediate code is generated, the interpreter converts the intermediate code to the target code. The target code is in JavaScript and follows p5.js' semantics to allow the output to be displayed on web browsers using its player. To this end, each entry in the \textsc{Drawing} dictionary is converted to a class. The entries in the \textsc{Animation} dictionary are processed to yield a series of conditionals, each corresponding to a sequence of frames of continuous, identical animation.

\lstdefinestyle{interfaces}{
  float=tp,
  floatplacement=tbp,
  abovecaptionskip=-5pt,
}

\lstset{
    language=Python, 
    commentstyle=\color{black}, keywordstyle=\color{black}, 
    stringstyle=\color{black}, 
    emphstyle={\color{black}},
    columns=fullflexible, 
    keepspaces=true, 
    breaklines=true, 
    frame=tb, 
    gobble=4, 
    breakindent=0pt, 
    belowcaptionskip=14pt, 
    basicstyle=\ttfamily, 
    mathescape, 
    numbers=left,
    stepnumber=1, 
    showstringspaces=false
    xleftmargin=4.4ex
}
\begin{lstlisting}[style = interfaces, label = intermediate-code, caption = {
    Intermediate Code Generated for Listing~\ref{code}}]
    "Drawing": {
      "nose": [{"fill": "#C2B280", 
        "stroke": "DEFAULT_STROKE", "shape": 
        "rectangle(-650, 250, 800, 15)"}],
      "pinocchio": [{"fill": "DEFAULT_FILL", 
        "stroke": "DEFAULT_STROKE", "shape": 
        "circle(0, 250, 200)"}, {"fill": "#C2B280", 
        "stroke": "DEFAULT_STROKE", "shape": 
        "rectangle(-650, 250, 800, 15)"}]
    }
    "Animation": {
      "nose": [{"action": "moveX", "start": 1, 
        "end": 101}, {"action": "moveX", "start": 102, 
        "end": 202}, {"action": "moveX", "start": 203, 
        "end": 303}]
    }
\end{lstlisting}

\vspace{5.1mm}
\subsection{Error Handling}

Separation of concerns is also observed in error handling. Lexical errors (e.g., a token not matching any of the patterns in the lexer grammar) are caught by the lexer during lexical analysis. Syntax errors (e.g., mismatched parentheses) are caught by the parser during syntactic analysis. To complement ANTLR's default error recovery strategy and to output more tailored error messages, the parser grammar also includes special production rules for catching common errors, such as specifying an incorrect number of coordinates in a \texttt{Coord} pair. Semantic errors (e.g., incompatible operands) are caught by the interpreter during semantic analysis.

The impact of error messages on the programming experience of novice programmers has been emphasized in human-computer interaction studies~\cite{10.1145/3441636.3442320, 10.1145/3555009.3555032, 10.1145/3411764.3445696}. To maximize the helpfulness and utility of error messages, AniFrame attempts to avoid terse, technical phrasing; for example, instead of "unsupported operand type(s) for \texttt{+}: \texttt{\textquotesingle int\textquotesingle} and \texttt{\textquotesingle str\textquotesingle}", AniFrame displays "\texttt{+} operator between \texttt{Number} and \texttt{Text} is not supported." Line numbers of errors are displayed, along with column numbers for lexical and syntactic errors. AniFrame also takes advantage of ANTLR's adaptive LL(*) parsing and error recovery strategy, which are purposely designed to reduce cascading error messages~\cite{10.5555/2501720, 10.1145/2714064.2660202}. 

\section{Sample Use Cases}
We demonstrate the utility and expressivity of AniFrame as a domain-specific language for creative coding through two sample use cases: animating a composite object and drawing a Sierpi\'{n}ski triangle via recursion. 

\subsection{Animating a Composite Object} \label{comp}
\begin{figure}[!t]
    \subfloat[Frame 1]{%
    \includegraphics[width=0.30\linewidth]{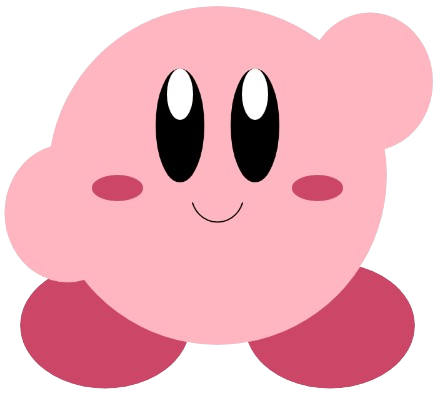} \label{kirby-frames1}}
    \subfloat[Frame 9]{%
   \includegraphics[width=0.31\linewidth]{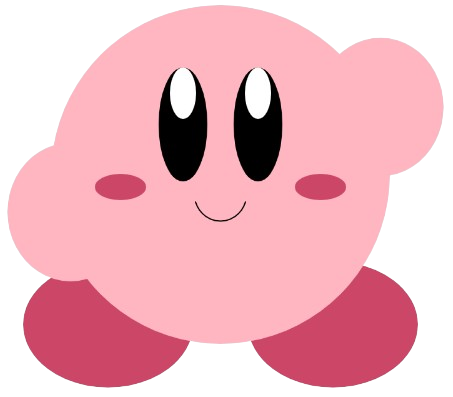} \label{kirby-frames2}}
    \subfloat[Frame 38]{%
   \includegraphics[width=0.32\linewidth]{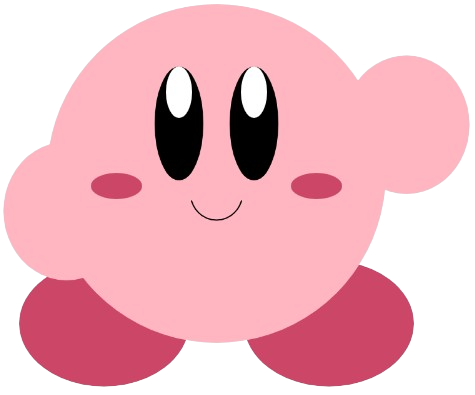} \label{kirby-frames3}}
    \caption{Composite Object Drawn and Animated in AniFrame. \textmd{The figure (a character sprite waving its left hand) shows selected frames from the output of the code in Listings~\ref{kirby} to \ref{ani_func}.}}
    \label{kirby-frames}
\end{figure}

The code for creating and animating a composite object in AniFrame and the output are provided in Listings~\ref{kirby} to \ref{ani_func} and Figure~\ref{kirby-frames}, respectively. As discussed in Section~\ref{data-types}, the semantics of the addition operator simplifies the layering of objects into composite objects (Lines 9, 23, and 37 in Listing~\ref{kirby}). It is also possible to create reusable object templates via user-defined functions (Listing~\ref{kirby_func} and Lines 17, 18, 20, and 21 in Listing~\ref{kirby}). 

The stacking of objects on the canvas is controlled by the order of the \texttt{draw()} statements in Lines~43 to 50 in Listing~\ref{kirby}; the parameters of these statements pertain to the start and end frames of the objects' appearance on the canvas. Meanwhile, Listing~\ref{ani_func} shows how custom animations can be created using AniFrame's built-in animation functions and control structures.

Other language constructs demonstrated in Listing~\ref{kirby} include explicit type declaration (Lines 7 and 8), type inferencing (Lines 1 and 2), and type coercion (Lines 1, 3, and 4; the inferred data type of \texttt{kabi\_color} is \texttt{Text}, but it was implicitly converted to \texttt{Color} when passed as an argument to \texttt{fill()} and \texttt{stroke()}).

\lstdefinestyle{interfaces}{
  float=tp,
  floatplacement=tbp,
  abovecaptionskip=-5pt,
}

\lstset{
    language=Python, 
    commentstyle=\color{black}, keywordstyle=\color{black}, 
    stringstyle=\color{black}, 
    otherkeywords={circle, fill, stroke, ellipse, draw, returns},
    emph={Object, Color, Coord, List, Text, Number},
    emphstyle={\color{black}},
    classoffset=1, % starting new class
    morekeywords={make_part, make_hand, },
    keywordstyle=\color{black},
    classoffset=0,
    columns=fullflexible, 
    keepspaces=true, 
    breaklines=true, 
    frame=tb, 
    gobble=4, 
    breakindent=0pt, 
    belowcaptionskip=14pt, 
    basicstyle=\ttfamily, 
    mathescape, 
    numbers=left,
    stepnumber=1, 
    xleftmargin=4.4ex
}
\begin{lstlisting}[style = interfaces, label = kirby, caption = {
    Sample Code for Creating a Composite Object}]
    kabi_color = "#ffb6c1"
    kabi_body = circle(250, 250, 270)
    kabi_body.fill(kabi_color)
    kabi_body.stroke(kabi_color)
    
    cheeks_color = "#cc4668"
    l_cheek: Object = ellipse(170, 260, 40, 20)
    r_cheek: Object = ellipse(330, 260, 40, 20)
    kabi_cheeks: Object = l_cheek + r_cheek
    kabi_cheeks.fill(cheeks_color)
    kabi_cheeks.stroke(cheeks_color)

    # NOTE: Insert function definition of make_part(), as given in Listing 4

    black = "#000000"
    white = "#ffffff"
    l_eye = make_part(220, 210, 38, 90, black)
    l_eye_shine = make_part(220, 185, 20, 40, white)
    l_eye += l_eye_shine 
    r_eye = make_part(280, 210, 38, 90, black)
    r_eye_shine = make_part(280, 185, 20, 40, white)
    r_eye += r_eye_shine
    kabi_eyes: Object = l_eye + r_eye
    
    smile: Object = curve(230, 150, 230, 272, 270, 272, 270, 150)
    smile.stroke(black)
    smile.fill(kabi_color)
    
    l_foot = make_part(160, 370, 135, 100, cheeks_color)
    r_foot = make_part(340, 370, 135, 100, cheeks_color)
    
    # NOTE: Insert function definition of make_hand(), as given in Listing 4

    r_hand = ellipse(130, 280, 100, 110)
    l_hand = make_hand()

    hoshi_no_kabi: Object = r_hand + kabi_body
    hoshi_no_kabi.fill(kabi_color)
    hoshi_no_kabi.stroke(kabi_color)

    # NOTE: Apply animation on l_hand (Listing 5)
    
    l_foot.draw(1, 1000)
    r_foot.draw(1, 1000)
    l_hand.draw(1, 1000)
    hoshi_no_kabi.draw(1, 1000)
    l_eye.draw(1, 1000)
    r_eye.draw(1, 1000)
    kabi_cheeks.draw(1, 1000)
    smile.draw(1, 1000)
\end{lstlisting}

\lstset{
    language=Python, 
    commentstyle=\color{black}, keywordstyle=\color{black}, 
    stringstyle=\color{black}, 
    otherkeywords={circle, fill, stroke, ellipse, draw, returns, func, rand_int, sqrt, repeat, move},
    emph={Object, Color, Coord, List, Text, Number},
    emphstyle={\color{black}},
    classoffset=1, % starting new class
    morekeywords={make_part, make_hand, wave_hand, wave_hand_up, sierpinski},
    keywordstyle=\color{black},
    classoffset=0,
    columns=fullflexible, 
    keepspaces=true, 
    breaklines=true, 
    frame=tb, 
    gobble=4, 
    breakindent=0pt, 
    belowcaptionskip=14pt, 
    basicstyle=\ttfamily, 
    mathescape, 
    numbers=left,
    stepnumber=1, 
    xleftmargin=4.4ex
}
\begin{lstlisting}[style = interfaces, label = kirby_func, caption = {
    User-Defined Drawing Functions in Listing~\ref{kirby}}]
    func make_part(x: Number, y: Number, w: Number,  h: Number, color: Color) returns Object:
        part: Object = ellipse(x, y, w, h)
        part.fill(color)
        part.stroke(color)
        return part

    func make_hand() returns Object:
        sugoi = ellipse(360, 160, 100, 110)
        sugoi.fill(kabi_color)
        sugoi.stroke(kabi_color)
        return sugoi 
\end{lstlisting}

\begin{lstlisting}[style = interfaces, label = ani_func, caption = {
    Animating the Left Hand (\texttt{l\_hand}) in Listing~\ref{kirby}}]
    func wave_hand(frame: Number, delta: Number):
        repeat(3):
            l_hand.move(1, 2, frame, frame + delta)

    func wave_hand_up(frame: Number, delta: Number):
        repeat(3):
            l_hand.move(-1, -2, frame, frame + delta)
    
    frame = 1
    delta = 12  
    repeat(3):
        wave_hand(frame, delta)
        wave_hand_up(frame + delta, delta)    
        frame += 2 * delta
\end{lstlisting}

\subsection{Drawing with Recursion}
To show the computational expressivity of AniFrame, we use it to programmatically generate a Sierpi\'{n}ski triangle, a fractal with the overall shape of an equilateral triangle. It is recursively constructed by connecting the midpoints of an equilateral triangle, removing the central triangle formed, and repeating these steps for the remaining (smaller) equilateral triangles. 

\begin{figure}[!t]
    \subfloat[Step = 2]{%
    \includegraphics[width=0.31\linewidth]{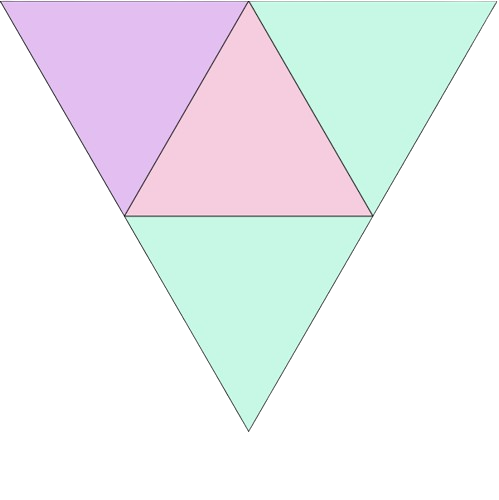} \label{6a}}
    \subfloat[Step = 3]{%
   \includegraphics[width=0.31\linewidth]{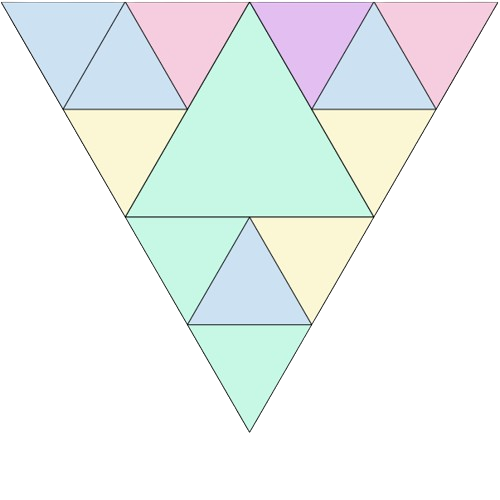} \label{6b}}
    \subfloat[Step = 8]{%
   \includegraphics[width=0.31\linewidth]{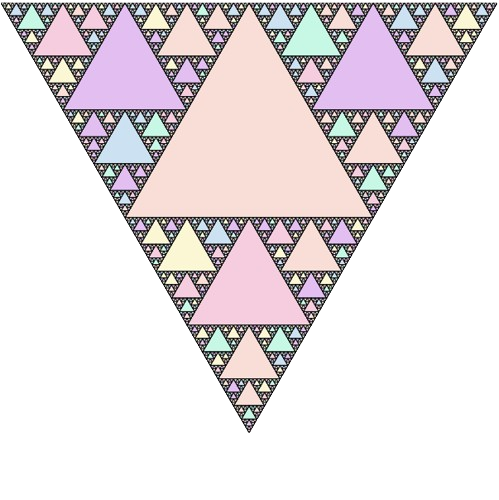} \label{6c}}
    \caption{Sierpi\'{n}ski Triangle Drawn in AniFrame. \textmd{The figure was generated programmatically using recursion. The code is given in Listing~\ref{sierpinski}; the step is dictated by the last argument in Line 14.}}
    \label{fig:siepinski}
\end{figure}

The code for creating a Sierpi\'{n}ski triangle in AniFrame and the resulting output are provided in Listing~\ref{sierpinski} and Figure~\ref{fig:siepinski}, respectively. Aside from recursion, this code also demonstrates other constructs in the language, including lists (Line~1 in Listing~\ref{sierpinski}), support for common mathematical operations (\texttt{rand\_int()} in Line~6 and \texttt{sqrt()} in Line~14), and global scoping for objects (Lines~7 and 13; global scoping is discussed in Section~\ref{scoping}).

\lstset{columns=fullflexible, keepspaces=true, breaklines=true, frame=tb, gobble=4, breakindent=0pt, belowcaptionskip=14pt, basicstyle=\ttfamily, mathescape, numbers=left,
stepnumber=1, xleftmargin=4.4ex}
\begin{lstlisting}[style = interfaces, label = sierpinski, caption = {
    Sample Code for Creating a Sierpi\'{n}ski Triangle}]
    colors = ["#CCE1F2", "#C6F8E5", "#FBF7D5",       "#F9DED7", "#F5CDDE", "#E2BEF1"]
    
    func sierpinski(x1: Number, y1: Number,         x2: Number, y2: Number, x3: Number, y3: Number, step: Number):
        if step != 0:
            shape = triangle(x1, y1, x2, y2, x3, y3)
            shape.fill(colors[rand_int(0, 5)])
            triangles += shape
            
            sierpinski(x1, y1, (x1+x2)/2, (y1+y2)/2, (x1+x3)/2, (y1+y3)/2, step-1)
            sierpinski(x2, y2, (x1+x2)/2, (y1+y2)/2, (x2+x3)/2, (y2+y3)/2, step-1)
            sierpinski(x3, y3, (x3+x2)/2, (y3+y2)/2, (x1+x3)/2, (y1+y3)/2, step-1)
    
    triangles: Object = point(0, 0)    
    sierpinski(0, 0, 600, 0, 300, 300*sqrt(3), 8)
    triangles.draw(1, 100)
\end{lstlisting}

\section{Usability Test}

\begin{table*}[t!]
  \caption{Tasks Given in the Usability Test. \textmd{U1, U2, and U3 had taken an introductory programming course but have been coding for less than four months. U4, U5, and U6 have at least one year of programming experience. A tick mark (\(\checkmark\)) indicates that the respondent was able to accomplish the task.}}
  \begin{tabularx}{\linewidth}{rXXcccccccccccc}
    \toprule
    \multirow{2}{*}{\#} & \multirow{2}{*}{Task} & \multirow{2}{*}{Intent Tested} & \multicolumn{6}{c}{AniFrame} & \multicolumn{6}{c}{p5.js} \\  
    \cline{4-15}
    & & & U1 & U2 & U3 & U4 & U5 & U6 & U1 & U2 & U3 & U4 & U5 & U6 \\
    \midrule 
    1 & Draw a circle and make it move from left to right & Drawing a shape and applying a single transformation & \(\checkmark\) & \(\checkmark\) & \(\checkmark\) & \(\checkmark\) & \(\checkmark\) & \(\checkmark\) &  \(\checkmark\) & \(\checkmark\) & \(\checkmark\) & \(\checkmark\) & \(\checkmark\) & \(\checkmark\) \\
    2 & Display the text "Hello world" on a blue canvas & Working with \texttt{Text}, drawing a text box, and setting canvas-level configurations & \(\checkmark\) & \(\checkmark\) & \(\checkmark\) & \(\checkmark\) & \(\checkmark\) & \(\checkmark\) & \(\checkmark\) & \(\checkmark\) & \(\checkmark\) & \(\checkmark\) & \(\checkmark\) & \(\checkmark\) \\
    3 & Draw a triangle and make it move up and down thrice at 5 frames per second & Drawing a simple shape and applying multiple transformations & \(\checkmark\) & \(\checkmark\) & & \(\checkmark\) & \(\checkmark\) &  \(\checkmark\) & & & & & \(\checkmark\) & \(\checkmark\) \\
    4 & Draw a traffic light by creating red, green, and yellow signal lights and enclosing them in a rectangular border & Working with \texttt{Color} and drawing a composite object & \(\checkmark\) & \(\checkmark\) & \(\checkmark\) & \(\checkmark\) & \(\checkmark\) & \(\checkmark\) & & \(\checkmark\) & \(\checkmark\) & \(\checkmark\) & \(\checkmark\) & \(\checkmark\) \\
    5 & Draw a simple face with eyes, nose, and mouth, and make the eyes (and only the eyes) move from left to right & Drawing a composite object and applying a transformation to only a component of it & \(\checkmark\) & \(\checkmark\) & & \(\checkmark\) & \(\checkmark\) & \(\checkmark\) & & & \(\checkmark\) & \(\checkmark\) & \(\checkmark\) & \(\checkmark\) \\
  \bottomrule
  \label{tasks}
\end{tabularx}
\end{table*}

In order to have an initial assessment of AniFrame's readability and writability, we conducted a preliminary usability test. Among the six respondents, three had taken an introductory programming course (covering variables, conditionals, and loops) but have been coding for less than four months; the other three have at least one year of programming experience.

The respondents were asked to complete five programmatic drawing and animation tasks (Table~\ref{tasks}) in both AniFrame and p5.js. All six respondents had no previous exposure to AniFrame and p5.js and instead referred to their respective documentation while doing the tasks; in order to mitigate the influence of the documentation, we patterned AniFrame's documentation after that of p5.js. The coding environment was a browser-based text editor without autocomplete. No time limit was set, but the respondents were allowed to give up if they felt that a task was too difficult.

As seen in Table~\ref{tasks}, all six respondents were able to complete Tasks~1 and 2 in both AniFrame and p5.js. Task~4 was completed by all six respondents in AniFrame and by five respondents in p5.js. On the other hand, Task~3 was accomplished by five respondents in AniFrame but only by two respondents in p5.js (both of which have at least one year of programming experience). Lastly, Task~5 was accomplished by five respondents in AniFrame and by four respondents in p5.js.

Task~3 involved making a triangle move up and down thrice. All six respondents first tried searching for a "move" function in the documentation but did not obtain any hits in p5.js since it uses "translate". In relation to this, U2, U4, and U5 mentioned that they appreciate AniFrame's use of the less technical term "move" since it is more accessible for beginners. 

Coding in p5.js, U2 and U4 were able to move the triangle unidirectionally via \texttt{translate()} and \texttt{Vector} but were unable to implement the required change in direction. U5 did not use these constructs and instead opted for a frame-by-frame approach, manually adjusting the \(y\)-coordinate of the triangle per frame (Listing~\ref{p4-p5}). U3 followed a similar approach but was unable to limit the motion to three times since they coded a loop for the counter whereas the semantics of p5.js require a conditional (Line~8 of Listing~\ref{p4-p5}).

Meanwhile, coding in AniFrame, all six respondents utilized the built-in \texttt{moveY()} function (Listing~\ref{p4-af}). The respondents found the task to be easier in AniFrame (U1, U2, U4, U5, and U6), noting that it led to "significantly more readable and fewer lines of code" (U1) and required "less mental gymnastics since the \texttt{repeat} loop and \texttt{moveY()} work as expected compared to p5" (U2). However, one of the respondents (U3) was unable to finish the task since they did not adjust the \texttt{frame} parameter of the \texttt{moveY()} statement that corresponds to the change in direction.

\lstset{
    language=Python, 
    commentstyle=\color{black}, keywordstyle=\color{black}, 
    stringstyle=\color{black}, 
    otherkeywords={circle, fill, stroke, ellipse, triangle, draw, returns, func, rand_int, sqrt, repeat, move, moveX, moveY, FRAME_RATE, CANVAS_WIDTH, CANVAS_BACKGROUND, CANVAS_HEIGHT},
    emph={Object, Color, Coord, List, Text, Number},
    emphstyle={\color{black}},
    classoffset=1, % starting new class
    morekeywords={make_part, make_hand, wave_hand, wave_hand_up, sierpinski},
    keywordstyle=\color{black},
    classoffset=0,
    columns=fullflexible, 
    keepspaces=true, 
    breaklines=true, 
    frame=tb, 
    gobble=4, 
    breakindent=0pt, 
    belowcaptionskip=14pt, 
    basicstyle=\ttfamily, 
    mathescape, 
    numbers=left,
    stepnumber=1, 
    xleftmargin=4.4ex
}
\begin{lstlisting}[style = interfaces, label = p4-p5, caption = {
    p5.js Code of U5 for Task 3 of Usability Test (Moving a Triangle Up and Down Thrice)}]
    function setup() {
        createCanvas(720, 400);   frameRate(5);
    }
    
    y = 75;   delta = 5;   ctr = 3;
    
    function draw() { 
        if (y <= 200 && y >= 0 && ctr + 1 >= 0) {
            background(200);   
            triangle(30, y, 58, y-55, 86, y);
            y += delta;
        } else {
            delta *= -1;   y += delta;   ctr--;
        }
    }
\end{lstlisting}

\begin{lstlisting}[style = interfaces, label = p4-af, caption = {
    AniFrame Code of U5 for Task 3 of Usability Test (Moving a Triangle Up and Down Thrice)}]
    set FRAME_RATE to 5
    i=0
    repeat(3): 
        shape = triangle(30, 75, 58, 20, 86, 75)
        shape.moveY(20, 1+i, 80+i)
        shape.moveY(-20, 81+i, 160+i)
        i += 160              
    shape.draw(1, 1000)
\end{lstlisting}

Task 5 involved drawing a simple face (with eyes, nose, and mouth) and animating only the eyes. Coding in p5.js, U2 was able to draw all the required facial features but, since they drew the eyes before the other features, the animation intended for the eyes cascaded down to the other features as a result of the stack-based semantics of p5.js (Section~\ref{frame}). This pain point was not observed in AniFrame since it was purposely designed such that an animation is applied only to the target object calling it. Moreover, the AniFrame code was described to be easier to write and trace since the target object is clearly specified when performing the animation function call (U2, U4, and U5).

From the post-task semi-structured interview and the respondents' impressions, all six respondents cited AniFrame's readability as a major contributing factor to its suitability for beginning programmers. Its smaller set of functions also made it "less intimidating and a good starting point for animation" (U6). 

With regard to the points for improvement, U3 and U5 noted that, although the frame-based approach allowed for fine-grained control, having to specify the start and end frames for every animation function call may not be readily intuitive for novice programmers and for those unfamiliar with the concept of frames in the first place. U2 suggested augmenting the coding environment with a visual representation of the frames vis-à-vis the animation timeline. Another recommendation was to allow for the canvas dimensions, canvas background, and frame rate to be accessible as special variables.

\section{Conclusion}
In this paper, we present AniFrame, an open-source domain-specific language for two-dimensional drawing and frame-based animation for novice programmers. Its design can be characterized as follows:

First, it features animation-specific data types, operations, and built-in functions for rapid creation and animation of composite objects. Second, it allows for fine-grained control over animation sequences through explicit specification of the target object, alongside the start and end frames. Third, it attempts to reduce the learning curve by adopting a Python-like syntax, supporting type inferencing, and using keywords that map closely to their semantic intent. Fourth, it promotes computational expressivity through built-in mathematical functions and support for recursion.

Our usability test points to AniFrame's potential to facilitate increased readability and writability for creative coding applications. Future directions include exploring syntactic and semantic improvements to enhance the intuitiveness of the frame-based approach, as well as providing support for three-dimensional graphics.

\bibliographystyle{ACM-Reference-Format}
\bibliography{sample-bibliography}

\end{document}